# PRIVACY PRESERVING MFI BASED SIMILARITY MEASURE FOR HIERARCHICAL DOCUMENT CLUSTERING

P. Rajesh[1], G. Narasimha[2], N.Saisumanth[3]

[1,3]Department of CSE, VVIT, Nambur, Andhra Pradesh, India
Email: rajesh.pleti@gmail.com
Email: saisumanth.nanduri@gmail.com

[2]Department of CSE, JNTUH, Hyderabad, Andhra Pradesh, India
Email: narasimha06@gmail.com

***Abstract:*** *The increasing nature of World Wide Web has imposed great challenges for researchers in improving the search efficiency over the internet. Now days web document clustering has become an important research topic to provide most relevant documents in huge volumes of results returned in response to a simple query. In this paper, first we proposed a novel approach, to precisely define clusters based on maximal frequent item set (MFI) by Apriori algorithm. Afterwards utilizing the same maximal frequent item set (MFI) based similarity measure for Hierarchical document clustering. By considering maximal frequent item sets, the dimensionality of document set is decreased. Secondly, providing privacy preserving of open web documents is to avoiding duplicate documents. There by we can protect the privacy of individual copy rights of documents. This can be achieved using equivalence relation.*

## I. INTRODUCTION

Document clustering has been studied intensively because of its wide applicability in areas such as web mining, search engines, text mining and information retrieval. The rapid progress of databases in every aspect of human actions has resulted in enormous demand for efficient algorithms for spinning data into valuable knowledge.

Document clustering has undergone through various methods, still document clustering is in its inefficiency state for providing the required information needed by the user exactly and approximately. Suppose the user makes an incorrect selection while browsing the documents in hierarchy. If user may not notice his mistakes until he browses into the deep portion of the hierarchy, then it decreases the efficiency of search and increases the number of navigation steps to find relevant documents. So we need a hierarchical clustering that is relatively flat that reduces the number of navigation steps. Therefore there is a great need for new document clustering algorithms, which are more efficient than conventional clustering algorithms [1, 2].

The increasing nature of World Wide Web has imposed great challenges for researchers to cluster the similar documents over the internet and their by improving the efficiency of search. Search engine uses the getting more confused in selecting the relevant documents among huge volumes of search results returned to a simple query. A potential solution to this problem is to cluster the similar web documents, which helps the user in identifying the relevant data easily and effectively [3].

The outline of this paper is divided into six sections. section II, briefly discusses related work. We explained our proposed algorithm description including common preprocessing steps and pseudo code of algorithm in section III. It also includes to precisely defining clusters based on maximal frequent item set (MFI) by Apriori algorithm. Section IV, describes exploiting the same maximal frequent item set (MFI) based similarity measure for Hierarchical document clustering with running example. In section V, provides privacy preserving of open web documents by using equivalence relation to protect the individual copy rights of a document.. Section VI, consists of conclusion and future scope.

## II. RELATED WORK

The related work of using maximal frequent item set in web document clustering is explained as follows. Ling Zhuang Honghua Dai [4] introduced a new criterion to specifically locate the initial points using maximal frequent item set. These initial points are then used as centers for k-means algorithm. However k-means clustering is completely unstructured approach, sensitive to noise and produces an unorganized collection of clusters that is not favorable to interpretation [5, 6]. To minimize the overlapping of documents, Beil, Ester [7] were proposed a method HFTC (Hierarchical Frequent Text Clustering) is another frequent item set based approach to choose the next frequent item sets. But the clustering result depends on the order of choosing next frequent item sets. The resulting hierarchy in HFTC usually contains many clusters at first level. As a result the documents in the same class are to be distributed into different branches of hierarchy, which decreases the overall clustering accuracy.

C.M.Fung [8] has introduced FIHC (Frequent Item set based Hierarchical Clustering) method for document clustering. Which employed, a cluster topic tree is constructed based on the similarity among clusters. FIHC used the efficient child pruning when number of clusters is large and to apply the elaborated sibling merging only when number of clusters is small. The experiment results FIHC actually outperforms all other algorithms (bisecting-k means, UPGMA) in accuracy for most number of clusters.

The Apriori algorithm [9] is a well-known method for computing frequent item sets in a transaction database. The document under the same topic, shares more common frequent item sets (terms) than the documents of different topics. The main advantage of using frequent item sets is that it can identify the relation among the more than two documents at a time in a document collection unlike similarity measure between two documents [10, 11].By the means of maximal frequent item sets, the dimensionality of the document set is reduced. More over maximal frequent item sets captures most related document sets. On the other hand, hierarchical clustering most relevant for browsing and maps most specific documents to generalized documents in the whole collection.

A conventional hierarchical clustering method constructs the hierarchy by subdividing parent cluster or merging similar children clusters. It usually suffers from its inability to perform tuning once a merge or split decision has been performed. This rigidity may lower the clustering accuracy. Furthermore, due to the fact that a parent cluster in the hierarchy always contains all objects of its Childs, this kind of hierarchy is not suitable for browsing. The user may have difficulty to locate his intention object in such a large cluster.

Our hierarchical clustering method is completely different. The aim of this paper is, first we form all the clusters by assigning documents to the most similar cluster using maximal frequent item sets by Apriori algorithm and then construct the hierarchical document clustering based on their inter-cluster similarities via same maximal frequent item set (MFI) based similarity measure . The clusters in the resulting hierarchy are non-overlapping. The parent cluster contains only the general documents.

### III. ALGORITHM DESCRIPTION

In this section, we explained our proposed algorithm description including common preprocessing steps and pseudo code of algorithm. It also includes to precisely defining clusters based on maximal frequent item set (MFI) by Apriori algorithm. First, we will speak about some common preprocessing steps for representing each document by item sets (terms). Second we will bring in vector space model by assigning weights to terms in all document sets. Finally, we will explain the process of initialization of clusters seeds using MFI to perform hierarchical clustering. Let Ds represents set of all documents in collection of database.

$$Ds= \{d1, d2, d3\ldots\ldots d_M\}: 1 \leq i \leq M$$

*A. Pre-Processing*

The document set Ds is converted from unstructured format into some common representation using the text preprocessing techniques, in which words or terms are extracted (tokenization). The input data set of documents in Ds are preprocessed using the techniques namely, removing HTML tags first, after that apply stop words list and stemming algorithm.

a) HTML Tags: parsing of HTML Tag
b) Stop words: Remove the stop words list like "conjunctions, connectives, prepositions etc"
c) Stemming algorithm: We utilize porter 2 stemmer algorithm in our approach.

*B. Vector representation of document:*

Vector space model is the most commonly used document representation model in text mining, web mining and information retrieval areas. In this model each document is represented as n-dimensional term vector. The value of each term in the n-dimensional vector reflects the importance of corresponding document. Let N be the total number of terms and M be the number of documents and each the document can be denoted as $D_i = (term_{i1}, term_{i2}, \ldots\ldots\ldots..term_{in})$ $1 \leq i \leq M$. Where $df(term_{ij}) < threshold$ value. The document frequency $term_{ij}$ is less than the threshold value is considered to avoid the problem of more times a term appears throughout all documents in the whole collection, the more poorly it discriminates between documents [12].Calculate term frequency tf is number of times a term appears in a document. Document frequency of a term df as no of documents that contains term. Also construct the weights for documents vectors. $D_i = (w_{i1}, w_{12}, w_{13}, \ldots\ldots, w1_{in})$

Where $w_{ij} = tf_{ij} * IDf(j)$ and

IDf (j) $= log\left(\frac{m}{df_j}\right) 1 \leq j \leq n$. where IDf is the inverse document frequency.

*Table 1: Table Representation of Transactional Database of Documents*

| Terms | Doc 1 | Doc 2 | Doc 3 | ..... | Doc 4 |
|---|---|---|---|---|---|
| Java | 1 | 1 | 0 | ..... | 1 |
| Beans | 0 | 1 | 0 | ..... | 0 |
| ..... | ..... | ..... | ..... | ..... | ..... |
| Servlets | 1 | 0 | 1 | ..... | 1 |

By the representation of document as vector form, we can easily identify which documents Contains the same features .The more features documents have in common, the more related they are. Thus, it is realistic to find well related documents. Assume that each document is an item in the transactional database; each term corresponds to a transaction. Our aim is to search for highly related documents "appearing" together with same features (the documents whose MFI features are closed). Similarly, the maximal frequent item set discovery in the transaction database serves the purpose of finding items of documents appearing together in many transactions. i.e., document sets which have large amount of feature in common.

### C. Apriori for maximal frequent item sets

Mining frequent item sets is a primary content of data mining that emphasizes particularly in finding the relation of different items in the large database. Mining frequent patterns is crucial problem in many data mining applications such as the discovery of association rules, correlations, multidimensional patterns, and other numerous important inferring patterns from consumer market basket analysis and web access etc. The association mining problem is formulated as follows: Given a large data base of set of items transactions, find all frequent item sets, where a frequent item set is one that occurs in at least a user-specified threshold value of the data base. Many of the proposed item set mining algorithms are a variant of Apriori, which employs a bottom-up, breadth first search that enumerates every single frequent item set. Apriori is a conventional algorithm that was first introduced] for mining association rules. Association can be viewed as two-step process as

(1) Identifying all frequent item sets

(2) Generating strong association rules from the frequent item sets

At first, candidate item sets are generated and afterwards frequent item sets are mined with the help of these candidate item sets. In the proposed approach, we have used only the frequent item sets for further processing so that, we undergone only the first step (generation of maximal frequent item sets) of the Apriori algorithm.

A frequent item set is a set of words which occurs frequently together and are good candidates for clusters and are denoted by FI. An item set X is closed if there does not exist an item set X1 such that X1, such that X ⊂ X1 and t(X) = t(X1), where t(X) defined as the set of transactions that contain item set X and it is denoted by FCI(frequently closed items).If X is frequent and no superset of X is frequent among the set of items I in transactional databases. Then we say that X is maximal frequent item set and denoted by MFI. Then MFI⊂ FCI ⊂ FI Whenever there are very long patterns are present in the data it is often impractical to generate the entire set if frequent item sets or closed item sets [16]. In that case, maximal frequent item sets are adequate for such applications. We employed maximal frequent item set algorithm from [17] using apriori. These maximal frequent item sets are initial seeds for hierarchical document clustering.

### D. Pseudo code Algorithm

For MFI Based Similarity Measure for Hierarchical Document Clustering

**Input:** Document set $D_s$.

**Definition:** MFI: Maximal Frequent Item set.

(tf) Term frequency and (df) document frequency

Step 1. For each document in $D_s$, Remove the HTML tags and perform stop word list and stemming.

Step 2. Calculate the term frequency (tf) and document frequency (df).

$D_i = (term_{i1}, term_{i2}, \ldots \ldots \ldots \ldots term_{in})$ 1≤i≤M

Where $df(term_{ij}) <$ Threshold value

Step 3. Also construct the weighted document vectors for all the documents

$D_i = (w_{i1}, w_{12}, w_{13}, \ldots \ldots, w1_{in})$ Where $w_{ij} = tf_{ij} * IDf(j)$.Idf (j) $= log\left(\frac{m}{df_j}\right)$ 1≤j≤n.

Step 4. Now represent each documents by keywords whose tf>support

Calculate the Maximal Frequent Item set(MFI) of terms using Apriori algorithm $MFI = \{F_1, F_2, F_3, \ldots \ldots \ldots \ldots F_n\}$

Where each $F_i = \{d_1, d_2, d_3, \ldots \ldots \ldots d_k\}$

Step 5. If a document $d_i$ is in more than one maximal frequent item set then choose $I_d$ as a set consisting of such maximal frequent item sets containing document $d_i$. Then Assign $I_x = I_{d0}$ .For each the maximal frequent item sets containing the document $d_i$

$If[jaccards(center (I_x, d_i))$
$> jaccards(center (I_{di}, d_i))]$

Then assign $I_x = I_{di}$. Assign the document $d_i$ to $I_x$ and discard $d_i$ for other maximal frequent item sets. Repeat this process for all documents that occurs in more than one maximal frequent item set

Step 6. Apply hierarchical document clustering to make these maximal frequent item sets $F_i$ as clusters and combine the documents in $F_i$ into a single new document and represent it by centers of the maximal frequent item sets. These are obtained by combining the features of maximal frequent item set of terms that grouping the documents

Step 7. Repeat the same process of hierarchical document clustering based on maximal frequent item sets for all levels in hierarchy and stop if total number of documents equals to one else go to step 4.

## IV. HIERARCHICAL CLUSTERS BASED ON MAXIMAL FREQUENT ITEM SETS

After finding maximal frequent item sets (MFI) by using Apriori algorithm. We turn to describing the creation of hierarchical document clustering using same similarity measure by MFI. A simple instance case of example is also provided to demonstrate the entire process. The set of maximal frequent item sets among the whole collection of documents $D_S$ by apriorialgorithm are $MFI = \{F_1, F_2, F_3 \ldots \ldots F_n\}$. Where each MFI consist of set of documents represented by $F_i = \{d_1, d_2, d_3 \ldots \ldots d_k\}$. Then consider total number of documents which occurs in maximal frequent item sets in MFI as follows.

$$MFI = \begin{Bmatrix} d_1, d_2, d_3, d_4, d_5, d_6, d_7, d_8, \\ d_9, d_{10}, d_{11}, d_{12}, d_{13}, d_{14}, d_{15} \end{Bmatrix}$$

$F_1 = \{d_2, d_4, d_6\}$

$F_2 = \{d_3, d_4, d_8\}$

$F_3 = \{d_1, d_5, d_7\}$

$F_4 = \{d_4, d_2, d_{14}\}$

$F_5 = \{d_{10}, d_{12}, d_{15}\}$

$F_6 = \{d_9, d_{11}, d_{13}\}$

The clusters in the resulting hierarchy are non-overlapping. This can be achieved through the following cases.

**Case1:** If $F_i, F_j$ are same then choose one in random to form cluster.

**Case2:** If $F_i, F_j$ are different then form clusters of documents contained in $F_i, F_j$ independently. In our example, the maximal frequent item set of documents in $F_3, F_5$ and $F_6$ are different. So we form a clusters according to the documents contained in $F_i$ like $F_3 = \{d_1, d_5, d_7\}$ as one cluster in hierarchy and represent it by center (as in step6).

**Case 3:** If $F_i, F_j$ contains some same documents among the documents list obtained from MFI. Let us consider the case of document $d_2$ is repeatedin more than one maximal frequent item sets$\{F_1 F_4\}$. Similarly $d_4$ is repeated in$\{F_1, F_2, F_4\}$. Then choose$I_d = \{F_1, F_2, F_4\} = \{I_{d0}, I_{d1}, I_{d2}\}$for document$d_4$. Assign $I_x = I_{d0} = F_1$. For each the maximal frequent item sets in $I_d$ containing the document $d_4$ from $I_{d0}$ to $I_{d2}$ calculate the measure

$$If[jaccards(center\ (I_x, d_4))\\ > jaccards(center\ (I_{di}, d_4))]$$

By using this jaccards measure, we can identify the document $d_4$ closest to which maximal frequent item set among maximal frequent item sets containing the document $d_4$. Then assign $I_x = I_{di}$.

Let's suppose that $d_4$ is closed to the maximal frequent item set $F_4$. Assign the document$d_4$to$I_x = I_{di} = F_4$and discard $d_4$ for other maximal frequent item sets. After this step, each document belongs to exactly one cluster. Similarly $d_2$ belongs to$F_1$. Repeat this process for all documents that occurs in more than one maximal frequent item set. Since the documents $d_2, d_4$ are repeated in$F_1, F_4$. The clusters that will form at the first level of hierarchy by applying step5 and step 6 are as follows.

$F_1 = \{d_2, d_6\}$

$F_2 = \{d_3, , d_8\}$

$F_3 = \{d_1, d_5, d_7\}$

$F_4 = \{d_4, , d_{14}\}$

$F_5 = \{d_{10}, d_{12}, d_{15}\}$

$F_6 = \{d_9, d_{11}, d_{13}\}$

The hierarchical diagram for the above form of maximal frequent item set clusters can be representing as follows. Repeat the same process of hierarchical document clustering based on maximal frequent item sets for all levels in hierarchy and stop if total number of documents equals to one else go to step 4.

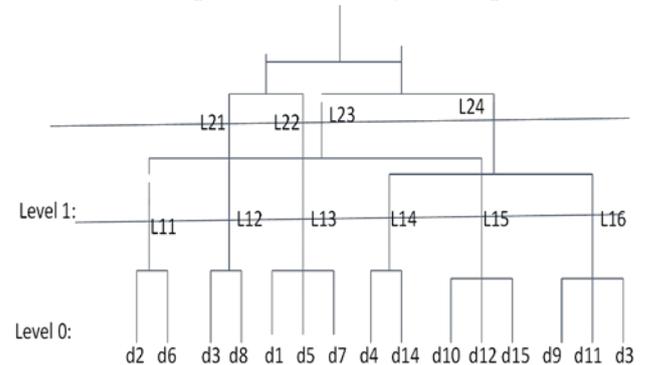

*Figure 1: Hierarchical document clustering using MFI*

Represent each new document $\{L_{ij}\}$ in hierarchy by maximal frequent item set of terms as centers (as in step 6). These maximal frequent item sets are obtained by combining the features of maximal frequent item set of terms that grouping the documents. Each new document also consisting of corresponding updated weights of maximal frequent item set of terms. Where $\{L_{ij}\}$ represents that $j^{th}$ document in the level of hierarchy $L_i$. In the figure $\{L_{12} = L_{21}\}$ means that the maximal frequent item set of terms in $2^{nd}$ document of level $L_1$ are not matched with other documents MFI set in same level $L_1$. So it is repeated same for the next level and it is also same for the document $\{L_{13} = L_{22}\}$. The documents $\{L_{11}, L_{15}\}$ and $\{L_{14}, L_{16}\}$ in first level are combined using MFI based hierarchical clustering and represent these documents in the second level as $L_{23}, L_{24}$.

## V. PRIVACY PRESERVING OF WEB DOCUMENTS USING EQUIVALENCE RELATION

Most internet web documents are publicly available for providing services required by the user. In such documents there is no confidential or sensitive data (open to all). Then how can we provide privacy of such documents. Now a days, same information will be exists in more than one document in duplicate forms. The way of providing privacy preserving of documents is by avoiding duplicate documents. There by we can protect the privacy of individual copy rights of documents. Many duplicate document detection techniques are available such as syntactic, URL based, semantic approaches. In each technique, a processing overhead of maintaining shingling's, signatures, fingerprints [13, 14, 15, 18]. In this paper, we proposed a new technique for avoiding duplicate documents using equivalence relation. Let Ds be the input duplicate document set is subset to web document collection. First find the jaccard similarity measure for every pair of documents in Ds using weighted feature representation of maximal frequent item sets discussed in step 2 and step 3 in algorithm. If the similarity measure of two documents is equal to 1, then the two documents are most similar. If the measure is 0, then they are not duplicates. The Jaccard index or the Jaccard similarity coefficient is a statistical measure of similarity between sample sets. For two sets, it is denoted as the cardinality of their intersection divided by the cardinality of their union. Mathematically

$$J(d_1, d_2) = \frac{|d_1 \cap d_2|}{|d_1 \cap d_2|}$$

For every pair of two documents calculate jaccard measure of d1, d2. All the diagonal elements in matrix are ones, because every document mostly related to itself. When we are classifying the documents into equivalence classes, we are not considering these ones and put zeros. Jaccard similarity coefficient matrix for four documents can be represented as follows.

$$R_\alpha = \begin{matrix} & d1 & d2 & d3 & d4 \\ d_1 & 1 & 0.4 & 0.8 & 0.5 \\ d_2 & 0.4 & 1 & 0.8 & 0.4 \\ d_3 & 0.8 & 0.8 & 1 & 0.9 \\ d_4 & 0.5 & 0.4 & 0.9 & 1 \end{matrix}$$

Where alpha is threshold. Let define a relation R on $D_s = \{d_1, d_2, d_3, d_4\}$ as the collection of document pairs whose similarity measure is above some threshold value. i.e $R = \{(d_i, d_j) / J(d_i, d_j) \geq threshold\}$

1. R is reflexive on Ds iff $R(d_i, d_i) = 1$. i.e Every document is mostly related to itself.

2. R is symmetric on Ds iff $R(d_i, d_j) = R(d_j, d_i)$ i.e if the document $d_i$ is similar to $d_j$ then the document $d_j$ is also similar to $d_i$.

3. R is transitive on Ds iff

   $R(d_i, d_k) \geq max_j \{ min\{R(d_i, d_j), R(d_j, d_i)\}\}$.

Then R is transitive by the definition.

Then R is an equivalence relation on Ds, which partitions the input document set Ds into set of equivalence classes. Equivalence relation seems a natural technique for duplicate document categorization. Any two documents in same equivalence class are related and are different if they are coming from two equivalence classes. The set of all equivalence classes induces the document set Ds. High syntactic similarity pairs of documents typically referred to as duplicates or near duplicates except diagonal elements. By using equivalence relation, easily we can identify the duplicate documents or we can perform the clustering on duplicate documents. Apart from the representation of feature document vector by MFI, we also need to consider that who is the author of document, when the document was created, where it is available, helps in effectively finding the duplicate documents. Each document in input Ds must belong to unique equivalence class. If R is equivalence relation on Ds = {d1, d2, d3, d4 …..d_n}. Then number of equivalence relations on Ds is always lies between $n \leq |R| \leq n^2$. i.e the time complexity of calculating equivalence relation on Ds is $O(n^2)$. Choose the threshold α in equivalence relation as 0.8 .i.e $J(d_i, d_j) \geq 0.8$. Since the matrix is symmetric, the documents sets $\{(d_3, d_1), (d_3, d_2), (d_4, d_3)\}$ are mostly related. Hence the documents are near duplicates and grouping the documents into clusters thereby providing privacy of individual copy rights of documents.

$$R_{0.8} = \begin{bmatrix} 0 & 0 & 1 & 0 \\ 0 & 0 & 1 & 0 \\ 1 & 1 & 0 & 1 \\ 0 & 0 & 1 & 0 \end{bmatrix}$$

## VI. CONCLUSION AND FUTURE SCOPE

Cluster analysis can be used as powerful ,stranded alone data mining concept that gains insight information of knowledge from huge unstructured databases. Most conventional clustering methods do not satisfy the document clustering requirements such as high dimensionality, huge volumes and easy of accessing meaningful clusters labels. In this paper, we presented novel approach; Maximal frequent item set (MFI) Based Similarity Measure for Hierarchical Document Clustering to address these issues. Dimensionality reduction can be achieved through MFI. By using the same MFI similarity measure in hierarchal document clustering, the number of levels will be decreased. It is easy for browsing. Clustering has its paths in many areas, by applying MFI based techniques to clusters, including data mining, statistics, biology, and machine learning we can get the high quality of clusters. Moreover, by means of maximal frequent item sets, we can predict the most influenced objects of clusters in the entire dataset of applications like business, marketing, world wide web, social networking analysis.

## VII. REFEERENCES